\documentclass[aps,prl,reprint,nofootinbib,twocolumn,superscriptaddress,showpacs,showkeys,longbibliography]{revtex4-1}
\usepackage{eurosym}
\usepackage{amsmath,amssymb,amstext}
\usepackage[usenames,dvipsnames]{color}
\usepackage{graphicx}
\usepackage{braket}
\usepackage{natbib}
\usepackage{comment}
\usepackage{dcolumn}
\usepackage[english]{babel}
\usepackage{wasysym}
\usepackage[colorlinks,bookmarks=false,citecolor=blue,linkcolor=red,urlcolor=blue]{hyperref}
\begin{document}
\title{Spinor polariton condensates under nonresonant pumping: Steady states and elementary excitations}
\author{Xingran Xu}
\affiliation{Shenyang National Laboratory for Materials Science, Institute of Metal Research, Chinese Academy of Sciences, Shenyang, China}
\affiliation{ School of Materials Science and Engineering, University of Science and Technology of China, Hefei, China}
\affiliation{Department of Physics, Zhejiang Normal University, Jinhua, 321004, China}
\author{Ying Hu}
\affiliation{State Key Laboratory of Quantum Optics and Quantum Optics Devices, Institute of Laser Spectroscopy,
Shanxi University, Taiyuan, Shanxi 030006, China}
\affiliation{Collaborative Innovation Center of Extreme Optics, Shanxi University,Taiyuan, Shanxi 030006, China}
\author{Zhidong Zhang}
\affiliation{Shenyang National Laboratory for Materials Science, Institute of Metal Research, Chinese Academy of Sciences, Shenyang, China}
\affiliation{ School of Materials Science and Engineering, University of Science and Technology of China, Hefei, China}
\author{Zhaoxin Liang}
\email{The corresponding author: zhxliang@gmail.com}
\affiliation{Department of Physics, Zhejiang Normal University, Jinhua, 321004, China}


\date{\today}

\begin{abstract}
We theoretically investigate a spinor polariton condensate under nonresonant pumping, based on driven-dissipative Gross-Pitaevskii equations coupled to the rate equation of a spin-unpolarized reservoir. We find the homogeneous polariton condensate can transit from the spin-unpolarized phase, where it is linearly polarized, to the spin-polarized phase, where it is elliptically polarized, depending on the cross-spin versus same-spin interactions and the linear polarization splitting. In both phases, we study elementary excitations using Bogoliubov approach, in a regime where the decay rate of total exciton density in reservoir crosses over from the slow to the fast limit. Depending on reservoir parameters, the global-phase mode can be either diffusive or gapped. By contrast, the relative-phase mode always possesses a gapped energy, undamped in the spin-unpolarized phase but weakly damped in the spin-polarized phase. In the spin-unpolarized phase, both modes are linearly polarized despite pumping and decay. However, in the spin-polarized phase, the mode polarization can be significantly affected by the reservoir and depends strongly on the circular polarization degree of the condensate. Interestingly, we demonstrate that the `ghost' branch of the Bogoliubov spectrum of the relative-phase mode can be visualized in the photoluminescence emission, distinguishable from that of the global-phase mode and thus allowing for experimental observation, when the spinor polariton condensate is elliptically polarized.
\end{abstract}

\maketitle

\section{Introduction}
At present, there are significant research interests in spinor exciton-polariton condensates in semiconductor microcavities \cite{Rev0,Rev1,Rev2,Rev3}. Formed from strong couplings between excitons and photons, polaritons possess peculiar spin properties \cite{Rev0}: the $J_z=\pm 1$ (spin-up or spin-down) spin projections of the total angular momentum of excitons along the growth axis of the structure directly correspond to the right- and left-circularly polarized photons absorbed or emitted by the cavity, respectively. The motivation behind the interests in spinor polariton condensates is two-fold. First,  a spinor polariton condensate is intrinsically nonequilibrium, with coherent and dissipative dynamics occurring on an equal footing \cite{Rev2,Sieberer2016}. This has resulted in numerous intriguing phenomena even in one-component polariton condensates \cite{Wouters2007,Oexcitation0,Byrnes2012,Xue2014,OneI1,OneI3,OneI4,OneI5,OneI7}. Further account of the polariton spin degree of freedom and their dynamics have revealed exceptionally rich physics in polariton systems \cite{Rev0}, such as the stimulated spin dynamics of polaritons \cite{SSDP1,SSDP2}, spin Meissner effect  \cite{Rubo2006,Larionov2010,Walker2011,Fischer2014}, optical spin Hall effect \cite{SHETheory1,SHEExperiment1,SHEExperiment2,SHEExperiment3}, spontaneous spin bifurcation \cite{Ohadi2015},  and ferromagnetic-antiferromagnetic phase transitions \cite{Ohadi2016,Ohadi2017} in spinor polariton Bose-Einstein condensate (BEC). Second, owing to the inherent spin multistabilities \cite{Spinint1,Spinint2} and fast spin dynamics, semiconductor microcavities bring prospects of implementing novel solid-state optoelectronic spin-logic architectures \cite{Liew2011}. First demonstration of polariton condensates as optical switches in state-of-the-art microcavity structure has been reported in recent experiments \cite{App1,App4}. Building on this theoretical and experimental progress, polariton-based systems \cite{Schneider2017} may promise a novel platform for realization and investigation of many-body systems.

Due to the spin structure of polaritons, the polariton-polariton interaction is strongly spin anisotropic \cite{Rev0}. In particular, the strength of the interaction constant $g_{12}$ for polaritons with opposite spin is typically much smaller than the interaction constant $g$ for same-spin polaritons \cite{Ciuti98, Attractive3}. As a result, the polariton condensate is generally expected to be linearly polarized \cite{Laussy2006,Shelykh2006} in the absence of mechanisms that \textit{explicitly} break the symmetry between the spin-up and -down components, such as magnetic field or spin-polarized pump. Still, a spontaneous magnetization of spinor polariton condensate has been observed which is induced by different loss rate of the linear polarizations \cite{Ohadi2015}.

Recently, several experiments have demonstrated the  possibility to tune the interaction constants using biexcitonic Feshbach resonance \cite{NP2014,Takemura2014,Feshbach2017} in resonantly created polariton condensate and single quantum well. Theoretically, this inspires an interesting question as regards the behavior of spinor polariton BEC formed under non-resonant pumping when the relative strength between cross-spin versus same-spin interactions can be varied in a wide range, although tuning interactions in this case remains experimentally challenging.

In this work, we theoretically investigate a spinor polariton BEC under nonresonant excitations in presence of linear polarization energy splitting denoted by $\Omega$, assuming tunability of the spin-anisotropic interactions. First, we study properties of a homogeneous polariton condensate having a uniform density $n_0$. We find that there exist two steady-state phases, a spin-unpolarized phase and a spin-polarized phase, depending on the parameter $\eta=g_{12}/[g+2\Omega/n_0]$: for $\eta<1$, the polariton condensate is in the spin-unpolarized phase, exhibiting a pinned linear polarization; for $\eta>1$, the condensate transits into the spin-polarized phase, where it becomes elliptically polarized. For the latter, whether the circular polarization is left or right handed is spontaneously chosen by the system. We note that our model assumes a spin-independent reservoir resulting from rapid spin relaxation, hence excludes the polarization transfer from the spin-polarized pump to the condensate as discussed in Refs. \cite{Liew2015,Li2015} using spin-polarized reservoir models. Moreover, the spontaneous creation of an elliptically polarized condensate in this work is induced by an interplay between interaction effects and linear polarization energy splitting, and occurs for $\eta>1$ which is beyond the typical experimental regime at present. This is different from Ref. \cite{Ohadi2015} in the regime $|g_{12}|\ll g$, where the elliptical polarization is induced by different energy and dissipation rates of the linear polarizations.

Second, we study elementary excitations in both phases of the spinor polariton condensate with the Bogoliubov-de Gennes (BdG) approach. Different from Refs. \cite{Shelykh2006,Rubo2006,SHELYKH2007313} in the context of equilibrium case and Ref. \cite{Solnyshkov2008} for resonantly created condensate, we study elementary excitations in a nonequilibrium polariton BEC, where the reservoir effect can significantly modify the energy and polarization of the collectively excited modes. Different from previous work (see, e.g., Ref. \cite{Borgh2010}) which assumes fast relaxation \cite{Bobrovska2015} of the incoherent reservoir density, here we consider effect of reservoir in the entire regime where the decay rate of the reservoir density crosses over from the slow to fast limit compared to the system dynamics.

We present detailed results on the energy spectrum and polarization of the global-phase mode and the relative-phase mode, corresponding to the global- and relative-phase excitations of the spinor components of the condensate, respectively. For energy spectra of excitations, we find that the global-phase mode is significantly affected by the reservoir. Depending on the reservoir parameters, the real part of the global-phase mode can be diffusive, gapped, even gapless. By contrast, the relative-phase mode always has a gapped real part of the energy spectrum, being undamped in the spin-unpolarized phase but weakly damped in the spin-polarized phase. For polarization of modes, we find that in the spin-unpolarized phase, both the global- and relative-phase modes are linearly polarized at all momenta, one copolarizing and the other cross polarizing with the linearly polarized condensate. This is similar as the equilibrium case \cite{Shelykh2006}, and is a consequence of symmetry properties of BdG equations in the spin-unpolarized phase regardless of effects of dissipation. However, different from the equilibrium case, the mode polarization in the spin-polarized phase can be significantly affected by reservoir particularly at low momenta, and how it varies with momenta also depends strongly on the circular polarization degree of the condensate.

Finally, we discuss how to probe the presented Bogoliubov dispersions of the spinor polariton BEC. Exploiting the photoluminescence (PL) emission, we show that both the negative-energy ghost dispersion branches of the spin mode and density mode can be directly observed in the PL spectrum, well distinguished from each other, when the polariton condensate is in the polarized phase.

The rest of the paper is structured as follows. In Sec.~\ref{section:one}, we present our theoretical model, based on which we solve for the stationary homogeneous state in Sec.~\ref{section:two}. There, we identify two phases of the spinor polariton BEC and discuss polarization properties of condensates. In Sec. ~\ref{section:three}, we present a comprehensive study on elementary excitations in both phases using the BdG theory, providing analytical and numerical results for the excitation spectrum and polarization properties of the excited modes. Then, in Sec.~\ref{section:four} we discuss how to experimentally probe the presented excitation spectrum exploiting PL emission. We conclude with a summary in Sec.~\ref{section:five}.

\section{Model}\label{section:one}

We consider a spinor exciton-polariton BEC created under non-resonant pumping in presence of linear polarization energy splitting. The order parameter for the condensate is described by a two-component time dependent wavefunction $\Psi=[\psi_1({\bf r},t),\psi_2({\bf r},t)]^T$, written in the basis of left- and right-circular polarized states \cite{Rev0}. For the excitonic reservoir, we assume that the spin relaxation of the reservoir is sufficiently fast so that the reservoir on the relevant time scales can be modeled by a scalar density denoted by $n_R(t)$. This way, we consider a situation when the reservoir does not explicitly affect the condensate polarization. Note that in realistic situations, the spin relaxation time of the reservoir is typically finite (see, e.g., Ref. \cite{TwoGP1}), which can be accounted with a reservoir model in terms of occupations of the left- and right-circular polarized components \cite{Liew2015,Li2015} rather than the total density.

The dynamics of the polariton condensate can be described by the driven-dissipative two-component Gross-Pitaevskii equation \cite{Borgh2010,Liew2015,Li2015,Askitopoulos2016}, i.e.,
\begin{eqnarray}
i\hbar\frac{\partial \psi_1}{\partial t}&=&\left[-\frac{\hbar^2\nabla^2}{2m}+g\left|\psi_1\right|^2+g_{12}\left|\psi_2\right|^2+g_Rn_R\right]\psi_1\nonumber\\
&-&\Omega\psi_2+\frac{i\hbar}{2}\left(R n_R-\gamma_C \right)\psi_1,\label{SpinorP1}\\
i\hbar\frac{\partial \psi_2}{\partial t}&=&\left[-\frac{\hbar^2\nabla^2}{2m}+g\left|\psi_2\right|^2+g_{12}\left|\psi_1\right|^2+g_Rn_R\right]\psi_2\nonumber\\
&-&\Omega\psi_1+\frac{i\hbar}{2}\left(R n_R-\gamma_C\right)\psi_2.\label{SpinorP2}
\end{eqnarray}
Here, $m$ is the mass of the polariton, $g$ ($g_{12})$ is the interaction constant between polaritons with same (opposite) spins, $\gamma_C$ is the decay rate of condensate polaritons, and $g_R$ characterizes the (spin-independent) interaction between the condensate and reservoir. The coherent spin flipping term $\Omega$ usually arises from the anisotropy-induced splitting of linear polarizations in the microcavity, as has been experimentally evidenced \cite{Klopotowski2006, Brunetti2007, Kasprzak2007}. For concreteness, we assume $\Omega>0$. We note that going beyond the Gross-Pitaevskii equations (\ref{SpinorP1}) and (\ref{SpinorP2}) to fully include the quantum and thermal fluctuations of the quantum field (e.g., Keldysh path-integral method \cite{ByPI1,ByPI2,ByPI0,ByPI01}) is beyond the scope of this work. Furthermore, in writing the above equations, we have assumed the situation where the transverse-electric and transverse-magnetic splitting \cite{Shelykh2004,Shelykh2006} vanishes.

We consider Eqs.~(\ref{SpinorP1}) and (\ref{SpinorP2}) are coupled to a (scalar) incoherent reservoir as mentioned earlier, described by a rate equation \cite{Pinsker2014}, i.e.,
\begin{equation}
\frac{\partial n_R}{\partial t}=P-\gamma_R n_R-R\left(\left|\psi_1\right|^2+\left|\psi_2\right|^2\right)n_R.\label{Rate}
\end{equation}
Here, $P$ is the rate of an off-resonant continuous-wave (cw) pumping, $\gamma_R^{-1}$ describes the life time of reservoir polaritons, and $R$ is the stimulated scattering rate of reservoir polaritons into the spinor condensate.

The properties of the spinor polariton BEC as stationary solutions to Eqs.~ (\ref{SpinorP1})-(\ref{Rate}) are determined by the rich interplay among the effects of pumping and decay, linear polarization energy splitting, and the spin-dependent interaction. In typical polariton systems, one has $g>0$, $g_{12}<0$, and $g\gg |g_{12}| $\cite{Rev0, Ciuti98, Attractive3}. In this case, polaritons driven by an off-resonant unpolarized pump tend to condense into a linearly polarized condensate \cite{Rev0,Read2009}. Its polarization direction is random for $\Omega=0$ as has been experimentally observed \cite{SPB1}, whereas a nonvanishing $\Omega$ will result in pinning of polarization \cite{Klopotowski2006, Kasprzak2007}. Inspired by recent experimental progress in realizing tunable cross-spin interaction properties \cite{NP2014,Feshbach2017}, below we are theoretically interested in the polariton BEC governed by Eqs.~(\ref{SpinorP1})-(\ref{Rate}) assuming $g_{12}/g$ can be flexibly controlled in a wide range, i.e., extends also into regimes that remain experimentally challenging to access.

\section{Homogeneous Steady States}\label{section:two}

Our goal is to seek the \textit{spatially homogeneous} stationary solutions to Eqs.~(\ref{SpinorP1})-(\ref{Rate}). We will use the pseudospin representation of the condensate because the condensate pseudospin vector $\vec{S}=\frac{1}{2}\left(\Psi^{\dagger}\cdot\sigma\cdot\Psi\right)$ with $\sigma_{x,y,z}$ the Pauli matrices provides an experimentally measurable quantity \cite{Rev0}. Substituting an ansatz of the form $\Psi_i({\bf r},t)=e^{-i\mu_T t}(\psi_{1}^0,\psi_{2}^0)^T=e^{-i\mu_T t}(\sqrt{S+S_z},\sqrt{S-S_z})^T$ and $n_R\left({\bf r},t\right)=n^0_R$ into Eqs.~(\ref{SpinorP1})-(\ref{Rate}), we obtain
\begin{eqnarray}
\dot{S}_{x}&=& -\left(\gamma_{C}-Rn_{R}\right)S_{x}+2\delta g S_{z}S_{y},\label{sx}\\
\dot{S}_{y}&=& -\left(\gamma_{C}-Rn_{R}\right)S_{y}-2\Omega S_{z}+2\delta g S_{z}S_{x},
\label{sy}\\
\dot{S}_{z}&=&-\left(\gamma_{C}-Rn_{R}\right)S_{z}-\Omega S_{y},\label{sz}\\
\dot{S}&=&-\left(\gamma_{C}-Rn_{R}\right)S,\label{s}\\
\dot{n}_{R}&=&P-\left\{ \gamma_{R}+2RS\right\} n_{R}. \label{n}
\end{eqnarray}
Here, we denote $\delta g=g-g_{12}$. Setting the left side of Eqs.~(\ref{sx})-(\ref{n}) to zero for stationary solutions, it follows from Eqs.~(\ref{s}) and (\ref{n}) that a condensate can spontaneously form, i.e, $S\neq 0$, if $P>P_{\textrm{th}}$ with $P_{\textrm{th}}=\gamma_R\gamma_C/R$ and $n^0_R=\gamma_C/R$. For condensate polarization, we see $S_y=0$ from Eq.~(\ref{sz}), but there exist two sets of stationary solutions for $S_x$ and $S_z$ according to Eq.~(\ref{sy}), i.e., $S_{z}\left(\Omega-\delta g S_{x}\right)=0$.

Depending on whether $S_z$ is zero or not in the stationary state, we identify two steady-state phases of the condensate, which we shall hereafter refer to as the spin-unpolarized phase and spin-polarized phase, respectively:

\begin{itemize}
\item{ Spin-unpolarized phase: $S_z=0$, $S_y=0$, and $S_x=S=n_0/2$, corresponding to an \textit{$X$-linearly polarized condensate} \cite{Rev0}, which exists under the condition $g_{12}<g+2\Omega/n_0$ with $n_0=\left(P-P_{\textrm{th}}\right)/(2\gamma_C)$. Correspondingly, we find $\mu_T=\left(g+g_{12}\right)n_0/2+g_Rn^0_R-\Omega$}. \\
\item{ Spin-polarized phase: $S_z\neq 0$, $S_y=0$, and $S_x=\Omega/(g_{12}-g)$, corresponding to an \textit{elliptically polarized condensate}, which exists  under the condition $g_{12}>g+2\Omega/n_0$. Moreover, we obtain $S_z=\pm (n_0/2)\sqrt{1-[2\Omega/(g-g_{12})n_0]^2}$, the sign being chosen randomly upon Bose condensation. Clearly, the circular polarization degree $s_z=|S_z/S|$ is given by $\sqrt{1-[2\Omega/(g-g_{12})n_0]^2}$. In addition, we find $\mu_T=gn_0+g_Rn^0_R$}.
\end{itemize}

We emphasize that, in our model, the form of Eqs.~(\ref{SpinorP1})-(\ref{Rate}) maintains the symmetry between the spin-up and -down polaritons, contrasting to Refs. \cite{Rubo2006,Liew2015,Li2015} where such symmetry is explicitly broken in some fashion. The transition from the spin-unpolarized to spin-polarized phases occurs at a critical interaction $g_{12}=g+2\Omega/n_0$. There, if a perturbation $\delta S_z$ is applied to the system Hamiltonian in form of $\lambda \sigma_z e^{(ikx-\omega t)+\eta t}$, with $\eta\rightarrow 0^+$, the linear response of the system, i.e., the spin density response function, can be obtained as $\chi_s \propto 1/(g-g_{12}+2\Omega/n_0)$, which diverges at the phase transition. We emphasize that while the spinor polariton condensate possesses a critical condition formally resembling the equilibrium counterpart \cite{Abad2013}, there is a fundamental difference due to the open-dissipative nature of our system where the condensate density $n_0$ is determined by the balance of pumping and decay.

\section{Elementary Excitations}\label{section:three}
\begin{figure*}
  \includegraphics[width=1.0\textwidth]{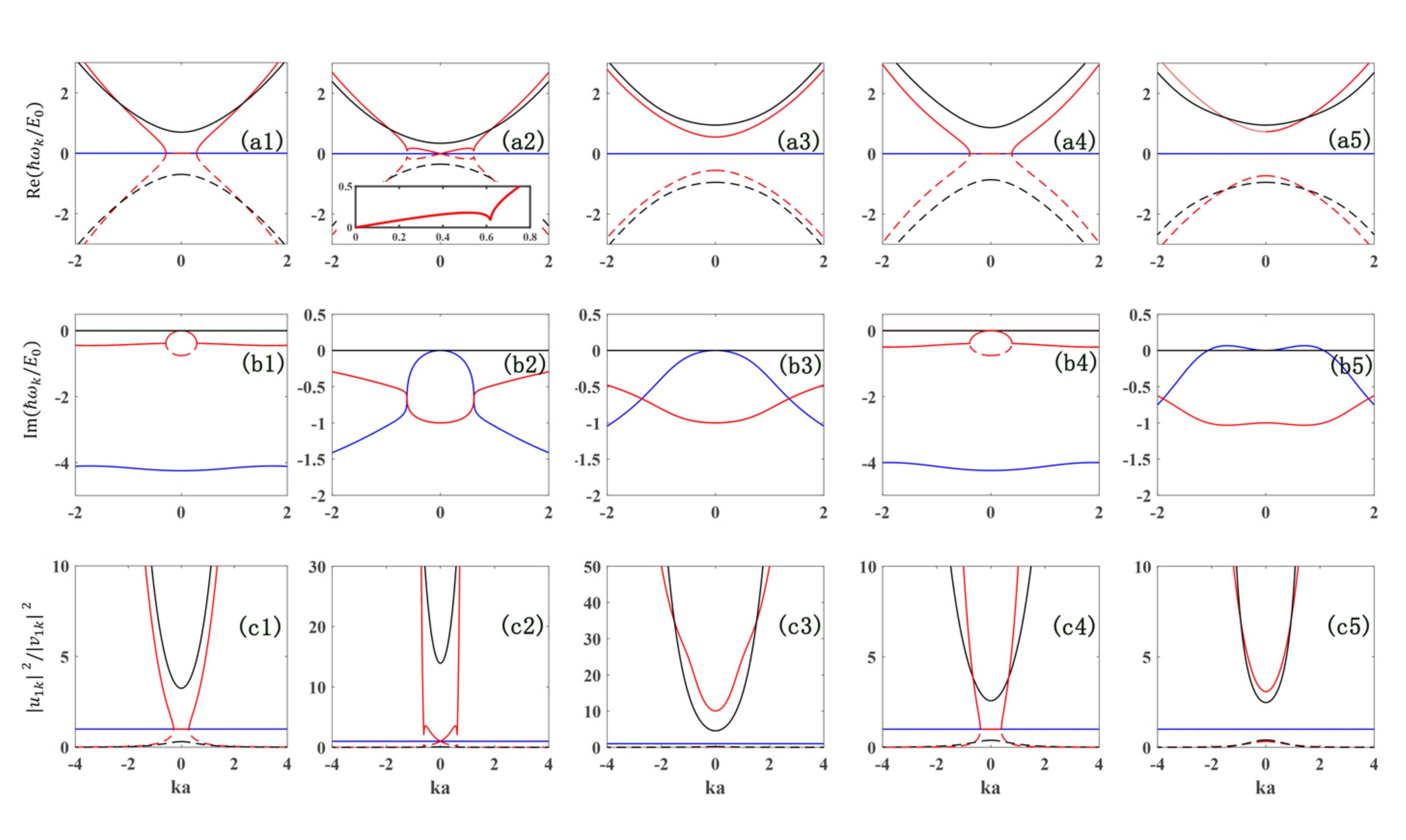}\\
\caption{ Elementary excitations of a linearly polarized polariton BEC. (a1)-(a5) Real part and (b1)-(b5) imaginary part of the excitation spectra associated with the global-phase modes (red curves), relative-phase modes (black curves), and the reservoir mode (blue curves), respectively; (c1)-(c5) the corresponding Bogoliubov amplitude ratio $\left|u_{1k}\right|^{2}/\left|v_{1k}\right|^{2}$ for the eigen-solutions of the matrix $\mathcal{L}_{{\bf k}}$ in Eq.~(\ref{BogMatrix1}). The solid and dashed curves in panels (a1)-(a5) depict the positive- and negative-energy spectrum of the excitation modes. In all plots, the length, the energy, and the time are scaled in units of $a=\sqrt{\hbar/m\gamma_C}$,  $E_0=\hbar \gamma_C$ and $\tau_c=\hbar/E_0$ with $a$ being the experimental length scale (e.g., in GaAs $a=2$ $\mu$m, $E_0=0.66$ meV, and $\tau_c=3$ ps), and we take $\Omega/E_0=0.1$. Other parameters [see Eq.~(\ref{BogMatrix1})] are chosen as follows: (a1)-(c1) $\gamma_R/\gamma_C=1.8$, $g_{12}/g=0.3$, and $P/E_0=5$; (a2)-(c2) $\gamma_R/\gamma_C=1$, $g_{12}/g=0.6$, and $P/E_0=2$; (a3)-(c3) $\gamma_R/\gamma_C=0.1$, $g_{12}/g=0.3$, and $P/E_0=2$;  (a4)-(c4) $\gamma_R/\gamma_C=1.8$, $g_{12}/g=-0.1$, and $P/E_0=5$; (a5)-(c5)  $\gamma_R/\gamma_C=0.1$, $g_{12}/g=-0.3$, and $P/E_0=2$. }\label{figure1}
\end{figure*}

The goal of this section is to investigate elementary excitations in the above two phases using the Bogoliubov approach \cite{Wouters2007,Liang0,Liang1,Liang2}. We start from the standard decomposition of the wave function $(\psi_1,\psi_2,n_R)^T$ into the steady-state solution $(\psi_1^0,\psi_2^0,n_R^0)^T$ and a small fluctuating term \cite{Wouters2007,Liang0,Liang1,Liang2}, i.e.,
\begin{widetext}
\begin{equation}\label{Bog1}
\begin{pmatrix}
\psi_{1}\left({\bf r}, t\right)\\
\psi_{2}\left({\bf r}, t\right)
\end{pmatrix}
= e^{-i\mu_T t}\begin{pmatrix}
\psi_{1}^0\\
\psi_{2}^0
\end{pmatrix}\left[
1
+\sum_{{\bf k}}\left\{
\begin{pmatrix}
u_{1{\bf k}}\\
u_{2{\bf k}}
\end{pmatrix}
e^{i\left({\bf k}{\bf r}-\omega t\right)}
+\begin{pmatrix}
v^*_{1{\bf k}}\\
v^*_{2{\bf k}}
\end{pmatrix}
e^{-i\left({\bf k}{\bf r}-\omega^* t\right)} \right\}\right],
\end{equation}
and
\begin{equation}
n_R\left({\bf r}, t\right)=n_R^0\left[1+\sum_k\left\{w_{\bf k}e^{i\left({\bf k}{\bf r}-\omega t\right)}+w^*_{\bf k}e^{-i\left({\bf k}{\bf r}-\omega^* t\right)}\right\}\right].\label{Bog2}
\end{equation}
Substituting Eqs.~(\ref{Bog1}) and (\ref{Bog2}) into Eqs.~(\ref{SpinorP1})-(\ref{Rate}) and retaining only first-order terms of fluctuation, we obtain at each momentum ${\bf k}$ the Bogoliubov-de Gennes (BdG) equation $\mathcal{L}_{\bf k}\mathcal{U}_{\bf k}=\hbar \omega_{{\bf k}} \mathcal{U}_{{\bf k}}$. Here, $\mathcal{U}_{\bf k}=\left(u_{1{\bf k}},v_{1{\bf k}},u_{2{\bf k}},v_{2{\bf k}},w_{\bf k}\right)^T$ and the operator $\mathcal{L}_{\bf k}$ in the matrix form reads as
\begin{equation}
\mathcal{L}_{{\bf k}}=
\left(\begin{array}{ccccc}
h_{1} & gn_1^{0}& g_{12}n_2^0-\Omega\sqrt{n^0_2/n^0_1} & g_{12}n_2^0 & g_{R}n_{R}^{0}+\frac{i}{2}Rn_{R}^{0}\\
-gn_1^{0} & -h_{1} & -g_{12}n_2^0 & -g_{12}n_2^0+\Omega\sqrt{n^0_2/n^0_1} & -g_{R}n_{R}^{0}+\frac{i}{2}Rn_{R}^{0}\\
g_{12}n_1^{0}-\Omega\sqrt{n^0_1/n^0_2} & g_{12}n_1^{0} & h_{2} & gn_2^{0} & g_{R}n_{R}^{0}+\frac{i}{2}Rn_{R}^{0}\\
-g_{12}n_1^{0} & -g_{12}n_1^{0}+\Omega\sqrt{n^0_1/n^0_2} & -gn_2^{0} & -h_{2} & -g_{R}n_{R}^{0}+\frac{i}{2}Rn_{R}^{0}\\
-iRn_1^{0} & -iRn_1^{0} & -iRn_2^{0} & -iRn_2^{0}& -i\left[Rn_{0}+\gamma_{R}\right]
\end{array}\right), \label{BogMatrix}
\end{equation}
\end{widetext}
with $h_{1(2)}=\varepsilon_{k}^{0}+g_{R}n_{R}^{0}+2gn_{1(2)}^{0}+g_{12}n_{2(1)}^{0}-\mu_T$, where $\varepsilon_{k}^{0}=\hbar^2k^2/2m$. Solutions to Eq.~(\ref{BogMatrix}) provide full specifications of the elementary excitations in the spinor polariton BEC.

As a consequence of dissipation, the Liouvillian matrix is non-Hermitian, and Eq.~(\ref{BogMatrix}) yields five complex dispersion branches: $\omega_j=\textnormal{Re}(\omega_j)+i\textnormal{Im}(\omega_j)$ ($j=1,2,3,4,5$), where the imaginary part represents the damping spectrum. Below, we present a detailed analysis on the energy of the Bogoliubov excitation modes, and their polarizations which may be accessed experimentally \cite{Rev0,Regemortel2015}, for the spin-unpolarized phase (see Sec.~\ref{Section:mis}) and spin-polarized phase (see Sec.~\ref{Section:in}), respectively. We will focus, in particular, on the different behavior of modes in two phases, and the effects of reservoir in the crossover regime from $\gamma_R\ll \gamma_C$ to $\gamma_R\gg \gamma_C$.

\subsection{Elementary excitations from the linearly polarized condensates}\label{Section:mis}

For the linearly polarized condensate formed in the regime $g_{12}<g+2\Omega/n_0$, we substitute the corresponding stationary values (see Sec.~\ref{section:one}) into the BdG Eq.~(\ref{BogMatrix}), giving
\begin{widetext}
\begin{equation}
\mathcal{L}_{{\bf k}}=
\left(\begin{array}{ccccc}
h_{1} & \frac{gn_{0}}{2} & \frac{g_{12}n_{0}}{2}-\Omega & \frac{g_{12}n_{0}}{2} & g_{R}n_{R}^{0}+\frac{i}{2}Rn_{R}^{0}\\
-\frac{gn_{0}}{2} & -h_{1} & -\frac{g_{12}n_{0}}{2} & -g_{12}n_{2}^{0}+\Omega & -g_{R}n_{R}^{0}+\frac{i}{2}Rn_{R}^{0}\\
\frac{g_{12}n_{0}}{2}-\Omega & \frac{g_{12}n_{0}}{2} & h_{2} & \frac{gn_{0}}{2} & g_{R}n_{R}^{0}+\frac{i}{2}Rn_{R}^{0}\\
-\frac{g_{12}n_{0}}{2} & -\frac{g_{12}n_{0}}{2}+\Omega & -\frac{gn_{0}}{2} & -h_{2} & -g_{R}n_{R}^{0}+\frac{i}{2}Rn_{R}^{0}\\
-iRn_{0}/2 & -iRn_{0}/2 & -iRn_{0}/2 & -iRn_{0}/2 & -i\left[Rn_{0}+\gamma_{R}\right]
\end{array}\right), \label{BogMatrix1}
\end{equation}
\end{widetext}
with $h_{1}=h_{2}=\varepsilon_{k}^{0}+gn_{0}/2+\Omega$.

Our main findings for elementary excitations in this case are as follows: (i) The global-phase mode is strongly affected by reservoir, such that its dispersion can be diffusive, gapped, or gapless, depending on $\gamma_C/\gamma_R$. By contrast, the relative-phase mode is always undamped with a gapped real energy; (ii) The global-phase mode copolarizes with the condensate while the relative-phase mode cross-polarizes with it, similar as that of the equilibrium condensate \cite{Shelykh2006} despite effects pumping and decaying. In the following, we will first present our results on the excitation spectrum for various reservoir parameters, before discussing polarization of collective modes.

\subsubsection{Excitation spectrum}

\textit{Two limiting cases.} - In the limit of vanishing reservoir when $n^0_R\approx 0$ and $R\approx 0$, by disregarding the reservoir effect in Eq.~(\ref{BogMatrix1}) in the leading order we find
\begin{equation}
\mathcal{L}_{{\bf k}}=
\left(\begin{array}{cccc}
h_{1} & \frac{gn_{0}}{2} & \frac{g_{12}n_{0}}{2}-\Omega & \frac{g_{12}n_{0}}{2}\\
-\frac{gn_{0}}{2} & -h_{1} & -\frac{g_{12}n_{0}}{2} & -\frac{g_{12}n_{0}}{2}+\Omega\\
\frac{g_{12}n_{0}}{2}-\Omega & \frac{g_{12}n_{0}}{2} & h_{2} & \frac{gn_{0}}{2}\\
-\frac{g_{12}n_{0}}{2} & -\frac{g_{12}n_{0}}{2}+\Omega & -\frac{gn_{0}}{2} & -h_{2}
\end{array}\right).\nonumber
\end{equation}
The Bogoliubov excitation spectra are found as
\begin{eqnarray}
\hbar\omega_{D}&=&\sqrt{\varepsilon_{k}^{0}\left[\varepsilon_{k}^{0}+\left(g+g_{12}\right)n_{0}\right]},\label{SBEC1} \\
\hbar\omega_{S}&=&\sqrt{\varepsilon_{k}^{0}\left(\varepsilon_{k}^{0}+\delta g n_{0}+4\Omega\right)+2\Omega\left(\delta g n_{0}+2\Omega\right)},\label{SBEC2}
\end{eqnarray}
where $\delta g=g-g_{12}$, and $\omega_D$ and $\omega_S$ correspond to energies of global-phase and relative-phase excitations, respectively. The global-phase mode is gapless at $k=0$, while the relative-phase mode exhibits an energy gap $\sqrt{2\Omega\left[\delta g n_0+2\Omega\right]}$ due to linear polariton splitting ($\Omega\neq 0$), which closes at the critical interaction strength $g_{12}=g+2\Omega/n_0$. For small momenta, the global-phase mode exhibits a linear dispersion while the relative-phase mode has an effective mass and a quadratic dispersion $\sim$$k^2$. We note that Eqs.~(\ref{SBEC1}) and (\ref{SBEC2}) are formally similar to the excitation spectra in an atomic coupled spinor BEC in equilibrium (see, e.g., Refs.  \cite{Abad2013,SpinorBEC1}), except that $n_0$ here is determined by the open-dissipative nature of polariton fluid.

In the opposite limit of fast reservoir, when $1/\gamma_R$ is the shortest time scale, an adiabatic elimination of the fast dynamics of reservoir in lowest order of the perturbation theory reduces Eq.~(\ref{BogMatrix1}) to the following matrix form:
\begin{widetext}
\begin{equation}
\mathcal{L}_{{\bf k}}=
\left(\begin{array}{cccc}
h_{1}-\frac{g\Gamma}{2R}-\left(\frac{i}{4}\Gamma+\omega\right) & \frac{gn_{0}}{2}-\frac{g\Gamma}{2R}-\frac{i}{4}\Gamma & \frac{g_{12}n_{0}}{2}-\frac{g\Gamma}{2R}-\Omega-\frac{i}{4}\Gamma & \frac{g_{12}n_{0}}{2}-\frac{g\Gamma} {2R}-\frac{i}{4}\Gamma\\
-\frac{gn_{0}}{2}+\frac{g\Gamma }{2R}-\frac{i}{4}\Gamma & -h_{1}+\frac{g\Gamma}{2R}-\left(\frac{i}{4}\Gamma+\omega\right) & -\frac{g_{12}n_{0}}{2}+\frac{g\Gamma }{2R}-\frac{i}{4}\Gamma & -g_{12}n_{2}^{0}+\frac{g\Gamma}{2R}+\Omega-\frac{i}{4}\Gamma\\
\frac{g_{12}n_{0}}{2}-\frac{g\Gamma}{2R}-\Omega-\frac{i}{4}\Gamma & \frac{g_{12}n_{0}}{2}-\frac{g\Gamma n_0}{2R}-\frac{i}{4}\Gamma & h_{2}-\frac{g\Gamma}{2R}-\left(\frac{i}{4}\Gamma+\omega\right) & \frac{gn_{0}}{2}-\frac{g\Gamma}{2R}-\frac{i}{4}\Gamma\\
-\frac{g_{12}n_{0}}{2}+\frac{g\Gamma}{2R}-\frac{i}{4}\Gamma & -\frac{g_{12}n_{0}}{2}+\frac{g\Gamma}{2R}+\Omega-\frac{i}{4}\Gamma & -\frac{gn_{0}}{2}+\frac{g\Gamma }{2R}-\frac{i}{4}\Gamma & -h_{2}+\frac{g\Gamma}{2R}-\left(\frac{i}{4}\Gamma+\omega\right)
\end{array}\right).\nonumber
\end{equation}
\end{widetext}
with $\Gamma=Rn_{0}\gamma_{C}/\left(\gamma_{R}+Rn_{0}\right)$. The resulting excitation spectra are found as
\begin{eqnarray}
\hbar\omega_{D}&=&-\frac{i\Gamma}{2}+\sqrt{\varepsilon_{k}^{0}\left[\varepsilon_{k}^{0}+\left(g+g_{12}\right)n_{0}\right]-\frac{\Gamma^{2}}{4}},\label{DBEC1} \\
\hbar\omega_{S}&=&\sqrt{\varepsilon_{k}^{0}\left(\varepsilon_{k}^{0}+\delta g n_{0}+4\Omega\right)+2\Omega\left[\delta g n_{0}+2\Omega\right]}. \label{DBEC2}
\end{eqnarray}
Comparisons of Eqs.~(\ref{SBEC2}) and  (\ref{DBEC2}) show $\omega_S(k)$ of the relative-phase mode stays the same in the two limits. However, $\omega_D(k)$ is strongly modified by the reservoir [see Eqs.~(\ref{SBEC1}) and (\ref{DBEC1})], with the low-lying global-phase mode transforming from the sound mode to diffusive mode \cite{Wouters2007}, whose energy is purely imaginary with the imaginary part behaves as $\sim$$k^2$  [see Eq.~(\ref{DBEC1})].

\textit{Generic case.}-- For arbitrary parameter $\gamma_R/\gamma_C$, solutions to Eq.~(\ref{BogMatrix1}) can be exactly cast into the following form:
\begin{widetext}
\begin{eqnarray}
\left[\left(\hbar\omega\right)^{2}-\left(\hbar\omega_{S}\right)^{2}\right]\times\left[\left(\hbar\omega\right)^{3}+i\left(Rn_{0}+\gamma_{R}\right)\left(\hbar\omega\right)^{2}-\left[Rn_{0}\gamma_{C}+\left(\hbar\omega_{D}\right)^{2}\right]\hbar\omega+ic(k)\right]=0. \label{Phase1General}
\end{eqnarray}
\end{widetext}
Here, $\omega_D$ and $\omega_S$ are given by Eqs.~(\ref{SBEC1}) and (\ref{SBEC2}), and
\begin{equation}
c(k)=-\left(Rn_{0}+\gamma_{R}\right)\left(\hbar\omega_{D}\right)^{2}+gn_{0}\gamma_{C}\varepsilon_k^0, \nonumber
\end{equation}
which tends to zero for $k\rightarrow0$. We immediately see from Eq.~({\ref{Phase1General}) that:

(1) There always exists two \textit{real} eigen-energy solutions $\omega=\pm\omega_S$ for the relative-phase modes, regardless of values of $\gamma_R/\gamma_C$, i.e., the relative-phase modes are not damped due to a decoupling from both the global-phase mode and reservoir modes. Indeed, as confirmed by the numerical results in Figs.~\ref{figure1}(a) and \ref{figure1}(b), the relative-phase modes (black curves) always exhibit a gapped real energy spectrum, and display qualitatively similar features despite variations of $\gamma_R/\gamma_C$. Although, the size of the gap can be tuned via variations of $g_{12}/g$ [c.f. Figs.~\ref{figure1}(a1) and \ref{figure1}(a2)], as expected from Eq.~(\ref{SBEC2}).

(2) By contrast, the global-phase modes and reservoir mode according to Eq.~(\ref{Phase1General}) display different behavior depending on $\gamma_R/\gamma_C$. In fact, decoupled from the relative-phase mode, the global-phase and reservoir modes are expected to exhibit similar properties as their counterpart in the one-component polariton condensate \cite{Wouters2007}, except for a modification by the interspecies coupling $g_{12}$ in $\omega_D$.At $k=0$, Eq.~(\ref{Phase1General}) becomes $\left(\hbar\omega\right)^{3}+i\left(Rn_{0}+\gamma_{R}\right)\left(\hbar\omega\right)^{2}-Rn_{0}\gamma_{C}\hbar\omega=0$, yielding three solutions: $\omega^{(0)}_{k=0}=0$ and
\begin{equation}
\omega^{\pm}_{k=0}=-i\left(\frac{Rn_{0}+\gamma_{R}}{2}\right)\pm\sqrt{Rn_{0}\gamma_{C}-\frac{1}{4}\left(Rn_{0}+\gamma_{R}\right)^2}.\nonumber
\end{equation}
Obviously, when $\gamma_{C}>(Rn_0+\gamma_{R})^2/(4Rn_0)$, $\omega^\pm_{k=0}$ have finite real components representing an energy gap, i.e., $E_\Delta=(1/2)\sqrt{4Rn_{0}\gamma_{C}-\left(Rn_{0}+\gamma_{R}\right)^2}$. In this case, $\omega^{\pm}_{k=0}$ correspond to gapped global-phase modes decaying at a rate $[Rn_{0}+\gamma_{R}]/2$. That the global-phase mode becomes massive has also been discussed in Ref. \cite{Byrnes2012} for the one-component polariton BEC taking into account the effect of the bottleneck polaritons. However, when $\gamma_{C}\le (Rn_0+\gamma_{R})^2/(4Rn_0)$, in particular, when $\gamma_R\gg \gamma_C\sim n_0R$, $\omega^{\pm}_{k=0}$ become purely imaginary with $\omega^+_{k=0}\approx -i3\gamma^2_C/(2\gamma_R)$ and ${\omega}^-_{k=0}\approx -i\gamma_R$. While $\omega^-_{k=0}$ corresponds to the fast decaying reservoir mode, both $\omega^+_{k=0}$ and $\omega^{(0)}_{k=0}$ are associated with the diffusive Goldstone modes.

At large momenta $k\gg a^{-1}$ ($a=\hbar/\sqrt{mgn_0}$ is the usual coherent length of BEC and we introduce $E_0=\hbar^2/ma^2$), Equation (\ref{Phase1General}) can be approximately solved as: $\omega^{(0)}_{ka\gg 1}\sim -i(Rn_0+\gamma_R)$ and $\omega^{\pm}_{ka\gg 1}\sim \pm \omega_D-ig_{R}n_{0}\gamma_{C}\hbar^{2}\epsilon_k^0/\omega_D^2$, the former corresponding to the reservoir mode while the latter for the global-phase modes. Since the imaginary part of $\omega^{\pm}_{k\rightarrow \infty}$ scales as $\sim$$1/k^2$ and thus vanishes at $k\rightarrow \infty$, we see that the global-phase modes with large momentum behave universally as free particles without being damped, independent of $\gamma_R$ and $\gamma_C$. Moreover, comparing the damping rate of the reservoir modes at $k=0$ and $k\rightarrow\infty$, we see that the reservoir modes exhibit similar damping rate when $\gamma_R\gg \gamma_C\sim Rn_0$, as opposed to the case $\gamma_R\ll \gamma_C\sim Rn_0$, where the damping rate of the reservoir mode becomes obviously $k$ dependent, increasing from $0$ (at $k=0$) to a value $\sim$$Rn_0$ (at $k\rightarrow \infty$). Thus, whereas the excitation modes almost decouple from each other in the fast reservoir limit, the density excitation significantly mixes with the reservoir if $\gamma_C>\gamma_R$ instead.

The above analysis is corroborated by the numerical results of excitation spectra for various parameters, as summarized in Fig.~\ref{figure1}. For the global-phase modes (see red curves), we observe the characteristic Goldstone branch when $\gamma_R\gg\gamma_C$ [see Figs.~\ref{figure1}(a1) and ~\ref{figure1}(a4)], which disappears for $\gamma_R\ll\gamma_C$ when an energy gap opens instead [see Figs.~\ref{figure1}(a3) and \ref{figure1}(a5)]. Physically, the existence of the Goldstone mode in presence of a fast decaying reservoir can also be understood from the following perspective: when $\gamma_R\gg\gamma_C$, the (fast) reservoir is able to adiabatically follow the slow rotation of the condensate phase across the sample, i.e., adiabatically follow the Goldstone mode. At large momenta, the global-phase modes are seen to exhibit $\textrm{Re}(\omega_k)\sim k^2$ and a suppressed damping for all parameters, consistent with earlier discussions. Interestingly, in the crossover regime $\gamma_R\sim \gamma_C$ where the reservoir effect strongly influences the global-phase excitation, we observe emergence of a dispersion (real part) that possesses a \textit{maxon-roton-like} character [see Fig.~\ref{figure1}(a2)], i.e., a softening of an excitation mode occurs at intermediate momentum.

\textit{Stability analysis.} For $\gamma_R\ll \gamma_C\sim Rn_0$ and $g_{12}<0$, a spatially homogeneous spinor polariton condensate can become dynamically unstable, due to an exponential growth of reservoir modulations with time: the reservoir mode shows $\textrm{Im}[\omega_k]>0$ at small momenta [see black curves in Fig.~\ref{figure1}(b5)]. Such dynamical instability disappears if $g_{12}>0$ is taken instead [see Fig.~\ref{figure1}(b3)]. To understand this, we seek the condition for a homogeneous spinor condensate to be stable in the considered regime by solving Eq.~(\ref{Phase1General}) at small momenta. Recall that in this case, the reservoir mode has $\omega=0$ at $k=0$, therefore, we expect $\omega(k)$ to be small at $k\rightarrow 0$. Retaining only the term linear in $\omega$ in Eq.~(\ref{Phase1General}), we find $\hbar\omega_{k\rightarrow 0}\approx ic(k)/(Rn_0\gamma_C+\omega_D^2)$, with $c(k)/E^2_0\approx -2(ka)^2[(1+g_{12}/g)(Rn_0+\gamma_R)-\gamma_C]$. Thus in order for the low-lying reservoir mode to be stable requires $c(k)\le 0$, giving the stability condition
\begin{equation}
\left(\frac{g+g_{12}}{g}\right)\frac{Rn_0+\gamma_R}{\gamma_c}>1.\nonumber
\end{equation}
Obviously, for the parameter regime $Rn_0\sim \gamma_C\gg \gamma_R$, the above criterion sustains for $g_{12}>0$ but is violated when $g_{12}<0$, explaining what we see in Fig.~\ref{figure1}(b5). For $g_{12}<0$, $c(k)$ can change its sign from positive to negative when the momentum increases to values larger than $k_R\neq 0$, where $k_R$ is determined by $c(k_R)=0$ giving $k_Ra\approx \sqrt{2[\gamma_C/(Rn_0+\gamma_R)-(g+g_{12})/g]}$, i.e., the stability condition is violated at momenta $0<k<k_R$ [see Fig.~\ref{figure1}(b5)], leading to growing perturbations.

\bigskip

\subsubsection{Polarization of quasiparticles}\label{QuasiLinear}

Previous studies on equilibrium linearly polarized condensates (see e.g., \cite{Shelykh2006}) have shown that both the global- and relative-phase modes are linearly polarized, one copolarizing and the other cross polarizing with the condensate. In the presence of pumping and decay, as mentioned earlier, the operator $\mathcal{L}_k$ governing the BdG equation for a non-equilibrium condensate becomes non-Hermitian and involves coupling to the reservoir excitations [see Eq.~(\ref{BogMatrix1})]. However, the symmetry properties of $\mathcal{L}_k$ matrix in the spin-unpolarized phase (see details in Appendix \ref{AppendixA}) dictate the following exact relations: for the global-phase mode, we have
\begin{eqnarray}
u_{1{\bf k}}=u_{2{\bf k}}, \hspace{2mm} v_{1{\bf k}}=v_{2{\bf k}}, \label{global}
\end{eqnarray}
and for the relative-phase mode, one has
\begin{equation}
u_{1{\bf k}}=-u_{2{\bf k}}, \hspace{2mm}v_{1{\bf k}}=-v_{2{\bf k}}.\label{relative}
\end{equation}
Thus, we conclude that for an $X$-linearly polarized open-dissipative condensate, the global-phase mode remains copolarized with the condensate while the relative-phase mode is cross poloarized with it. For later reference, here we also present the analytical expressions for the Bogoliubov coefficients from solving Eq.~(\ref{BogMatrix1}). For the global-phase mode, we obtain
\begin{eqnarray}
u_{1{\bf k}}&=&\frac{ \hbar\omega_{D}+\left[\left(g+g_{12}\right)n_{0}+\epsilon_k^0\right]}{\sqrt{8\hbar\omega_{D}\left(\epsilon_k^0+\left(g+g_{12}\right)n_{0}\right)}}.\nonumber\\
v_{1{\bf k}}&=&\frac{\hbar\omega_{D}-\left[\epsilon_k^0+\left(g+g_{12}\right)n_{0}\right]}{\sqrt{8\hbar\omega_{D}\left(\epsilon_k^0+\left(g+g_{12}\right)n_{0}\right)}}, \nonumber
\end{eqnarray}
and for the relative-phase mode, we find
\begin{eqnarray}
u_{1{\bf k}}&=&\frac{\left(g+g_{12}\right)n_{0}/2}{\sqrt{\frac{\left(g+g_{12}\right)^{2}n_{0}^{2}}{2}-2\left[\epsilon_k^0+\frac{n_{0}}{2}\left(g-g_{12}\right)+2\Omega-\hbar\omega_{S}\right]^{2}}},\nonumber\\
v_{1{\bf k}}&=&\frac{\hbar\omega_{S}-\left[\epsilon_k^0+\frac{n_{0}(g-g_{12})}{2}+2\Omega\right]}{\sqrt{\frac{\left(g+g_{12}\right)^{2}n_{0}^{2}}{2}-2\left[\epsilon_k^0+\frac{n_{0}(g-g_{12})}{2}+2\Omega-\hbar\omega_{S}\right]^{2}}}.\nonumber
\end{eqnarray}

As $|u_{1{\bf k}}|=|u_{2{\bf k}}|$ and $|v_{1{\bf k}}|=|v_{2{\bf k}}|$ apply for both modes, we plot $|u_{1k}|^2/|v_{1k}|^2$ in Figs.~\ref{figure1}(c1)-\ref{figure1}(c5) for various reservoir parameters. For $\gamma_R\gg \gamma_C$, we see that the eigenvectors of the global-phase mode show the usual infrared divergence $v_{1{\bf k}}\rightarrow k^{-1/2}$ and $u_{1{\bf k}}\rightarrow -v_{1{\bf k}}$ at $k\rightarrow 0$, giving rise to $|u_{1{\bf k}}/v_{1{\bf k}}|=1$ at $k=0$. By contrast, for $\gamma_R/\gamma_C\ll 1$ when the global-phase modes become gapped, we see that $|u_{1{\bf k}}/v_{1{\bf k}}|>1$ for momenta $k\rightarrow 0$ [see red curves in Figs.~\ref{figure1}(c3) and \ref{figure1}(c5)]. Similar behavior is also observed in the plot of $|u_{2{\bf k}}/v_{2{\bf k}}|$ for the gapped relative-phase mode at small momenta [see black curves in Fig.~\ref{figure1}(c)].

\subsection{Elementary excitations from the elliptically polarized condensates}\label{Section:in}

\begin{figure*}
  \includegraphics[width=1.0\textwidth]{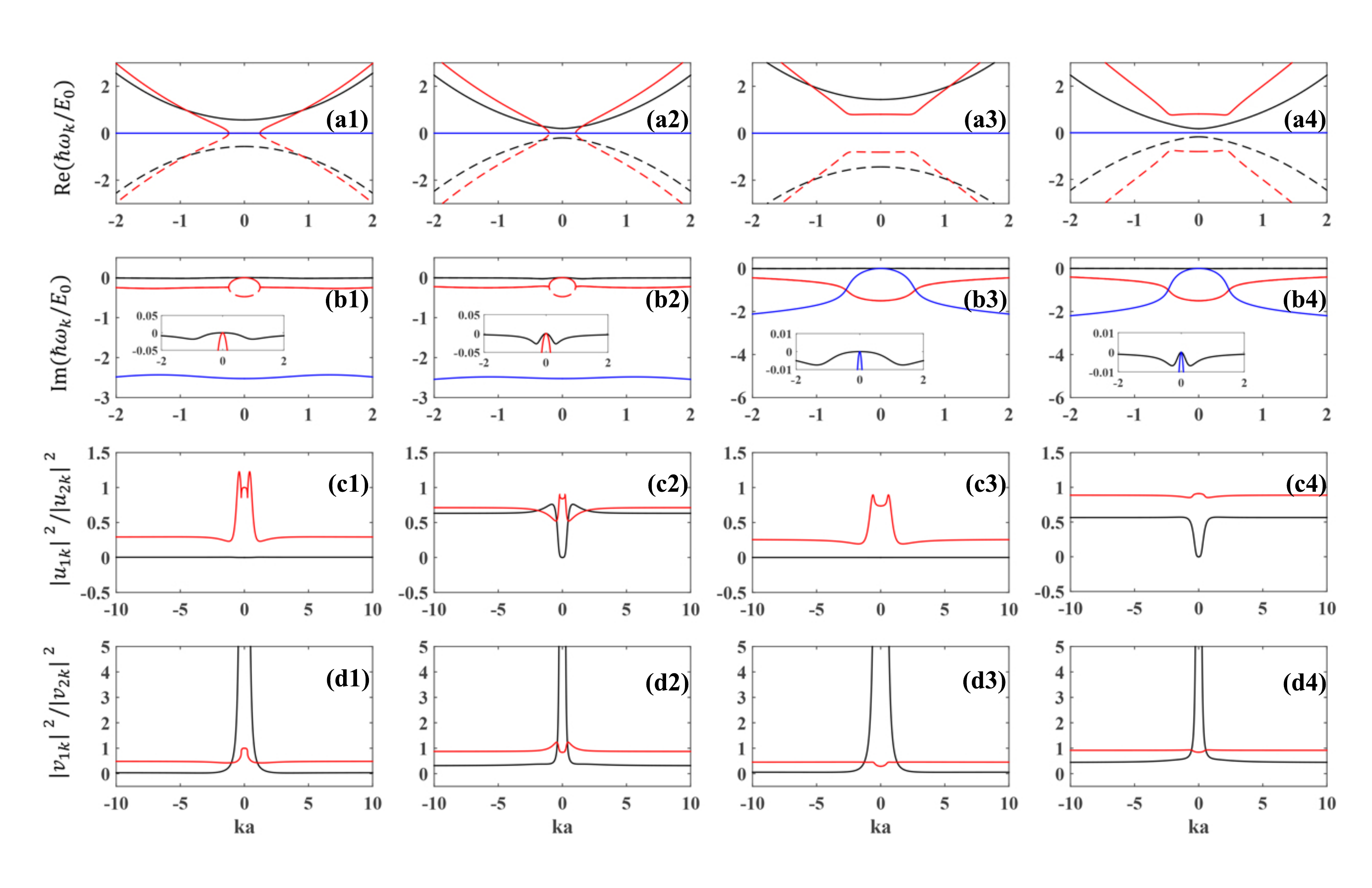}\\
\caption{ Elementary excitations in elliptically polarized polariton BEC. (a1)-(a4) Real part and (b1)-(b4) imaginary part of the excitation spectra associated with the global-phase modes (red curves), relative-phase modes (black curves), and reservoir modes (blue curves); ratios of Bogoliubov amplitudes (c1)-(c4) $\left|u_{1{\bf k}}\right|^{2}/\left|u_{2{\bf k}}\right|^{2}$, and (d1)-(d4) $\left|v_{1{\bf k}}\right|^{2}/\left|v_{2{\bf k}}\right|^{2}$ for eigen-solutions of the matrix $\mathcal{L}_{{\bf k}}$ in Eq.~(\ref{BogMatrix}). For parameters in Eq.~(\ref{BogMatrix}), we take  $P/E_0=3$, $R/E_0=1$ and (a1)-(d1) $\gamma_R/\gamma_C=1.8$, $g_{12}/g=1.5$, and $\Omega/E_0=0.1$; (a2)-(d2) $\gamma_R/\gamma_C=1.8$, $g_{12}/g=1.85$, and $\Omega/E_0=0.5$; (a3)-(d3) $\gamma_R/\gamma_C=0.1$, $g_{12}/g=1.5$, and $\Omega/E_0=0.1$; (a4)-(d4) $\gamma_R/\gamma_C=0.1$,  $g_{12}/g=1.35$, and $\Omega/E_0=0.5$. In all plots, the length, the energy and the time are scaled in units of $a=\sqrt{\hbar/m\gamma_C}$,  $E_0=\hbar \gamma_C$, and $\tau_c=\hbar/E_0$ with $a$ being the experimental length scale (e.g., in GaAs $a=2$ $\mu$m, $E_0=0.66$ meV and $\tau_c=3$ ps). Solid and dashed curves correspond to normal and ghost branches of excitations, respectively. }\label{figure2}
\end{figure*}

We now turn to the spin-polarized phase where the condensate is elliptically polarized in the regime $g_{12}>g+2\Omega/n_0$. The BdG Eq.~(\ref{BogMatrix}) now takes the form (we choose $S_z>0$ in our calculation)
\begin{widetext}
\begin{equation}
\mathcal{L}_{{\bf k}}=
\left(\begin{array}{ccccc}
h_{1} & \frac{gn_0\left(1+\Delta\right)}{2}& \frac{g_{12}n_0(1-\Delta)}{2}-\Omega\sqrt{\frac{1-\Delta}{1+\Delta}} & \frac{g_{12}n_0(1-\Delta)}{2} & g_{R}n_{R}^{0}+\frac{i}{2}Rn_{R}^{0}\\
-\frac{gn_0\left(1+\Delta\right)}{2} &-h_{1} & -\frac{g_{12}n_0(1-\Delta)}{2} & -\frac{g_{12}n_0(1-\Delta)}{2}+\Omega\sqrt{\frac{1-\Delta}{1+\Delta}}& -g_{R}n_{R}^{0}+\frac{i}{2}Rn_{R}^{0}\\
 \frac{g_{12}n_0(1-\Delta)}{2}-\Omega\sqrt{\frac{1+\Delta}{1-\Delta}} & \frac{g_{12}n_0(1-\Delta)}{2}  & h_{2} & \frac{gn_0\left(1-\Delta\right)}{2} & g_{R}n_{R}^{0}+\frac{i}{2}Rn_{R}^{0}\\
-\frac{g_{12}n_0(1-\Delta)}{2} & -\frac{g_{12}n_0(1-\Delta)}{2}+\Omega\sqrt{\frac{1+\Delta}{1-\Delta}} & -\frac{gn_0\left(1-\Delta\right)}{2} & -h_{2} & -g_{R}n_{R}^{0}+\frac{i}{2}Rn_{R}^{0}\\
-i\frac{n_0}{2}\left(1+\Delta\right)R & -i\frac{n_0}{2}\left(1+\Delta\right)R & -i\frac{n_0}{2}\left(1-\Delta\right)R & -i\frac{n_0}{2}\left(1-\Delta\right)R& -i\left[Rn_{0}+\gamma_{R}\right]
\end{array}\right), \label{BogMatrixPhase2}
\end{equation}
\end{widetext}
with $\Delta=\sqrt{1-[2\Omega/(g-g_{12})n_0]^2}$, $h_1=\varepsilon_{k}^{0}+gn_0\Delta+g_{12}n_0(1-\Delta)/2$, and $h_2=\varepsilon_{k}^{0}-gn_0\Delta+g_{12}n_0(1+\Delta)/2$.

Compared to the spin-unpolarized phase, our main findings on the elementary excitations from an elliptically polarized condensate are as follows: (i) The relative-phase mode becomes weakly damped. (ii) Polarization properties of the modes  are determined by interplay among the parameter $\gamma_R/\gamma_C$ of reservoir, momentum $k$ of the mode, and the circular polarization degree $s_z=2S_z/n_0$ of the condensate. Following, we present a detailed analysis on energy spectrum and polarization of modes in regimes of fast and slow reservoirs, respectively, as well as for elliptical polariton condensates with different circular polarization degree $s_z$.

\bigskip

\subsubsection{Excitation spectrum}

Figures \ref{figure2}(a) and \ref{figure2}(b) illustrate the excitation spectra for various parameters. Different from the spin-unpolarized phase, the relative-phase modes in the spin-polarized phases show a very weak damping [see black curves in Fig.~\ref{figure2} (b) and insets], along with a gapped (real part) energy spectrum. In addition, the global-phase modes and the reservoir modes exhibit similar features as that of the spin-unpolarized phase.

To gain more understanding on the excitation spectra illustrated in Fig.~\ref{figure2} (a) and \ref{figure2}(b), consider first the limit of vanishing reservoir, where Eq.~(\ref{BogMatrixPhase2}) reduces to
\begin{widetext}
\begin{eqnarray}
\mathcal{L}_{{\bf k}}=
\left(\begin{array}{ccccc}
h_{1} & \frac{gn_0\left(1+\Delta\right)}{2}& \frac{g_{12}n_0(1-\Delta)}{2}-\Omega\sqrt{\frac{1-\Delta}{1+\Delta}} & \frac{g_{12}n_0(1-\Delta)}{2} \\
-\frac{gn_0\left(1+\Delta\right)}{2} & -h_{1} & -\frac{g_{12}n_0(1-\Delta)}{2} & -\frac{g_{12}n_0(1-\Delta)}{2}+\Omega\sqrt{\frac{1-\Delta}{1+\Delta}}\\
 \frac{g_{12}n_0(1-\Delta)}{2}-\Omega\sqrt{\frac{1+\Delta}{1-\Delta}} & \frac{g_{12}n_0(1-\Delta)}{2}  & h_{2} & \frac{gn_0\left(1-\Delta\right)}{2} \\
-\frac{g_{12}n_0(1-\Delta)}{2} & -\frac{g_{12}n_0(1-\Delta)}{2}+\Omega\sqrt{\frac{1+\Delta}{1-\Delta}} & -\frac{gn_0\left(1-\Delta\right)}{2} & -h_{2}
\end{array}\right). \nonumber
\end{eqnarray}
\end{widetext}
Solutions of the eigen-energies are found as:
\begin{widetext}
\begin{eqnarray}
\left(\hbar\omega_{S/D}\right)^2&=&\varepsilon_{k}^{0}\left(\varepsilon_{k}^{0}+g_{12}n_{0}\right)+\frac{1}{2}\beta-2\Omega^{2}\nonumber\\
&\pm&\sqrt{\left[(\varepsilon_{k}^{0}(g_{12}-2g)n_{0}+\frac{1}{2}\beta\right]^{2}+\Omega^{2}\left[\frac{\hbar^{4}k^{4}(3g-g_{12})}{m\left(g_{12}-g\right)}+2\frac{\hbar^{2}k^{2}}{m}n_{0}(2g-g_{12})-2\beta\right]+4\Omega^{4}}, \label{Phase21}
\end{eqnarray}
\end{widetext}
where $\beta=(g-g_{12})^{2}n_{0}^{2}$ and, as before, $\hbar\omega_S$ and $\hbar\omega_D$ represent energies of the relative-phase mode and the global-phase mode, respectively. As in the spin-unpolarized phase, the global-phase mode is gapless while the spin mode has an energy gap $\omega_S(k)\rightarrow \sqrt{(g-g_{12})^{2}n_{0}^{2}-4\Omega^2}$.

In the opposite limit $\gamma_R/ \gamma_C\gg 1$ [see Figs.~\ref{figure2}(a1) and \ref{figure2}(a2), \ref{figure2}(b1) and \ref{figure2}(b2)], Equation (\ref{BogMatrixPhase2}) at the lowest order of $1/\gamma_R$ reads as:
\begin{widetext}
\begin{equation}
\mathcal{L}_{{\bf k}}=
\left(\begin{array}{cccc}
h_{1} & \tilde{g}n_{1}^{0} & \tilde{g}_{12}n_{2}^{0}-\Omega\sqrt{\frac{n_{2}^{0}}{n_{1}^{0}}} & \tilde{g}_{12}n_{2}^{0}\\
-\tilde{g}^{*}n_{1}^{0} & -h_{1}^{*} & -\tilde{g}_{12}^{*}n_{2}^{0} & -\tilde{g}_{12}^{*}n_{2}^{0}+\Omega\sqrt{\frac{n_{2}^{0}}{n_{1}^{0}}}\\
\tilde{g}_{12}n_{1}^{0}-\Omega\sqrt{\frac{n_{1}^{0}}{n_{2}^{0}}} & \tilde{g}_{12}n_{1}^{0} & h_{2} & \tilde{g}n_{2}^{0}\\
-\tilde{g}_{12}^{*}n_{1}^{0} & -\tilde{g}_{12}^{*}n_{1}^{0}+\Omega\sqrt{\frac{n_{1}^{0}}{n_{2}^{0}}} & -\tilde{g}^{*}n_{2}^{0} & -h_{2}^{*}
\end{array}\right),\nonumber
\end{equation}
with $\tilde{g}=g-\gamma(g+\frac{i}{2}R)$, $\tilde{g}_{12}=g_{12}-\gamma(g+\frac{i}{2}R)$, $\gamma=\gamma_{C}/(n_{0}\ensuremath{R}+\gamma_{R})$, $h_{1}=\varepsilon_{k}^{0}+g_{12}n_{2}^{0}+\left[g-\gamma(g+\frac{i}{2}R)\right]n_{1}^{0}-gn_{2}^{0}$, and $h_{2}=\varepsilon_{k}^{0}+g_{12}n_{1}^{0}+\left[g-\gamma\left(g+\frac{i}{2}R\right)\right]n_{2}^{0}-gn_{1}^{0}$. The eigen-energy is solution to the following equation:
\begin{eqnarray}
\left[\left(\hbar\omega\right)^{2}-a^{2}\right]\times\left[\left(\hbar\omega-ib\right)^{2}-c^{2}\right]+d=0,\label{Phase22}
\end{eqnarray}
with
\begin{eqnarray}
b&=&-\frac{1}{2}n_{0}R\gamma,\nonumber\\
a^{2}	&=&\left[\varepsilon_{k}^{0}+(g_{12}-g)n_{0}\right]\left[\varepsilon_{k}^{0}+(g_{12}-g)\left(n_{1}^{0}-n_{2}^{0}\right)\right],\nonumber\\
c^{2}	&=&-gn_{0}\varepsilon_{k}^{0}\gamma+(\varepsilon_{k}^{0})^2-\frac{1}{4}n_{0}^{2}R^{2}\gamma^{2}+\varepsilon_{k}^{0}(g_{12}-g)\left(n_{1}^{0}-n_{2}^{0}\right)+\varepsilon_{k}^{0}n_{0}(g+g_{12}),\nonumber\\
d&=&-4\varepsilon_{k}^{0}\Omega^{2}\left[n_{0}(g_{12}-g)+\varepsilon_{k}^{0}\right]\frac{n_{1}^{0}-n_{2}^{0}}{n_{0}}.\nonumber
\end{eqnarray}
\end{widetext}
At $k=0$, Eq.~(\ref{Phase22}) allows easy solutions: (i) solutions $\omega_{s}=\pm \sqrt{(g_{12}-g)^2n_0^2-4\Omega^2}$ correspond to the energy gap of the relative-phase modes [see black curves in Figs.~\ref{figure2}(a1) and \ref{figure2}(a2), \ref{figure2}(b1) and \ref{figure2}(b2)]; (ii) solutions $\omega_D^{(1)}=0$ and $\omega_D^{(2)}=-in_0R\Gamma/2$ are associated with the global phase excitation at $k=0$, near which the imaginary part scales as $\sim$$ k^2$ [see red curves in Fig.~\ref{figure2}(b1) and \ref{figure2}(b2)], indicating the existence of diffusive Goldstone mode [see also red curves in Figs.~\ref{figure2}(a1) and \ref{figure2}(a2)]. Noticing that the energy gap of the relative-phase mode is formally the same as the equilibrium counterpart [see Eq.~(\ref{Phase21}) at $k=0$], we anticipate the real part of the energy of the relative-phase mode is qualitatively similar for all reservoir parameters as in the linearly polarized condensate, consistent with what we see in Fig.~\ref{figure2}(a).

For arbitrary value of $\gamma_R/\gamma_C$, one can show that Eq.~(\ref{BogMatrixPhase2}) yields the following equation:
\begin{widetext}
\begin{eqnarray}
\left[\left(\hbar\omega\right)^{2}-\left(\hbar\omega_{D}\right)^{2}\right]\left[\left(\hbar\omega\right)^{2}-\left(\hbar\omega_{S}\right)^{2}\right]\Big[i\left(Rn_{0}+\gamma_{R}\right)+\hbar\omega\Big]+i\left(n_{0}\gamma_{C}\right){\left(g\frac{\hbar^2k^{2}}{m}+iR\left(\hbar\omega\right)\right)\left[\left(\hbar\omega\right)^{2}+F(k)\right]}=0,\label{eq:generic}
\end{eqnarray}
\end{widetext}
Here, $F(k)=4\Omega^{2}-\left[n_{0}(g_{12}-g)+\epsilon_k^0\right]^{2}+2\Omega^{2}k^{2}/(g_{12}-g)$, and $\omega_{S/D}$ are defined in Eq.~(\ref{Phase21}). Importantly, Equation (\ref{eq:generic}) suggests that the relative-phase mode solutions are no longer decoupled from the other modes, and, therefore, a damping in the relative-phase mode is generically expected [see black curves in Fig.~\ref{figure2}(b) and insets]. In addition, at $k=0$, Eq.~(\ref{eq:generic}) admits solutions: $\omega=0$, $\omega=-i(Rn_0+\gamma_R)/2\pm\sqrt{Rn_0\gamma_C-(Rn_0+\gamma_R)^2/4}$, and $\omega=\pm\omega_S$ (evaluated at $k=0$). Indeed, the energy gap of the relative-phase modes stays the same without modification from the reservoir, as confirmed by our numerical results in Fig.~\ref{figure2}(a). Moreover, a diffusive mode exists as long as $\gamma_C\le (Rn_0+\gamma_R)^2/(4Rn_0)$ holds, whereas an energy gap opens in the spectrum of the density excitation for $\gamma_C>(Rn_0+\gamma_R)^2/(4Rn_0)$, agreeing with the results in Fig.~\ref{figure2}(a) (see red curves).

\subsubsection{Polarization of quasiparticles}

We have shown in Sec.~\ref{QuasiLinear} that, in the spin-unpolarized phase, the global- and relative-phase modes are copolarized and cross polarized with the condensate, respectively, just as the equilibrium counterpart. As we will show, situations are significantly different for the spin-polarized phase. In analyzing mode polarization, we are interested in the quantities $|u_{1k}|^2/|u_{2k}|^2$ and $|v_{1k}|^2/|v_{2k}|^2$, as motivated by Ref. \cite{Regemortel2015}. There, it is proposed that the degree of polarization of the collective modes can be partially probed through the measurement of normal polar polarization angle of the eigenvector defined by $\cos(2\theta)=(|u_{1k}|^2-|u_{2k}|^2)/(|u_{1k}|^2+|u_{2k}|^2)$, or equivalently, $|u_{1k}|^2/|u_{2k}|^2=[\cos(2\theta)+1]/[\cos(2\theta)-1]$. In Figs.~\ref{figure2}(c) and \ref{figure2}(d), we plot $|u_{1k}|^2/|u_{2k}|^2$ and $|v_{1k}|^2/|v_{2k}|^2$ as functions of momenta, respectively, for both the global- and relative-phase modes (only those with positive real energies are shown). We note, however, that detailed discussions on the measurement of polarization of quasiparticles of the spinor polariton condensate are beyond the scope of this paper.

To see how the reservoir affects the polarization of modes in general, we first consider a strongly circularly polarized condensate ($s_z\approx 0.9$), and compare polarization of $[u_{1k}, u_{2k}]^T$ of both modes in the regime of a fast reservoir [see Fig.~\ref{figure2}(c1)] and a slow reservoir  [see Fig.~\ref{figure2}(c3)], respectively. In both regimes, the relative-phase mode (black curves) is seen to be circularly polarized at all momenta with a circular polarization opposite to the condensate. By contrast, $[u_{1k}, u_{2k}]^T$ of the global-phase mode (red curves) is elliptically polarized, whose polarization direction in particular at small momenta is strongly influenced by the reservoir: it exhibits a dramatic, even irregular, variation near $k=0$ in the fast reservoir regime [see Fig.~\ref{figure2}(c1)], as opposed to a more regular and smooth behavior in the slow reservoir case [see Fig.~\ref{figure2}(c3)]. This can be understood by noticing the different energy spectra in the two reservoir regimes: the global-phase mode exhibits diffusive dispersion in the limit of fast reservoir, whereas being gapped for a slow reservoir. At large momenta where the reservoir effect becomes unimportant, the global-phase mode in both plots exhibits similar polarization properties.

Similar features as above also appear for the global-phase modes of the condensate with $s_z\approx 0.1$, i.e., only a small asymmetry exists between the spin-up and spin-down components, as illustrated in Figs.~\ref{figure2}(c2) and \ref{figure2}(c4). In particular, $[u_{1k}, u_{2k}]^T$ of the global-phase mode in the fast reservoir regime [see red curve in Fig.~\ref{figure2}(c2)] exhibits rich variations near $k=0$, contrasting to a more flattened behavior in the slow regime [see Fig.~\ref{figure2}(c2)]. However, $[u_{1k}, u_{2k}]^T$ of the relative-phase mode is significantly affected by $s_z$. When the condensate is highly circular ($s_z\sim 1$),  $|u_{1k}|^2/ |u_{2k}|^2\approx 0$ at all momenta [see black curve in Fig.~\ref{figure2}(c1)], i.e., $[u_{1k}, u_{2k}]^T$ always has a strong strong circular polarization which is opposite to the condensate. Instead,  for $s_z\ll 1$ counterpart [see black curve in Fig.~\ref{figure2}(c2)], $[u_{1k}, u_{2k}]^T$ is strongly circular at $k=0$ and becomes elliptical away from it, with some strong variation near $k=0$. In the limit of large momenta, $|u_{1k}|^2/|u_{2k}|^2$ saturates to a constant smaller than $1$, which corresponds to an elliptical polarization whose circular polarization is opposite to the condensate.

Compared to $[u_{1k}, u_{2k}]^T$ in Fig.~\ref{figure2}(c), we see that the circular polarization of $[v_{1k}, v_{2k}]^T$ of the relative-phase mode is always opposite to $[u_{1k}, u_{2k}]^T$ at $k=0$ (same as the condensate), but is same as $[u_{1k}, u_{2k}]^T$ at large momenta, as illustrated by the black curves in Figs.~\ref{figure2}(d1)-\ref{figure2}(d4). For strongly circularly polarized condensate [see Figs.~\ref{figure2}(d1) and \ref{figure2}(d3)], $[v_{1k}, v_{2k}]^T$ is highly circularly polarized at all momenta, flipping its circular polarization rapidly from the left to the right when $k$ changes from $k=0$. Such flip is more steep when the condensate has $s_z\ll 1$ [see black curves Figs.~\ref{figure2}(d2) and \ref{figure2}(d4)]. In this case, $[v_{1k}, v_{2k}]^T$ of the relative-phase mode is elliptically polarized at large momenta. For the global phase mode (red curves), the corresponding $[v_{1k}, v_{2k}]^T$ is always elliptically polarized, displaying a more uniform behavior in the slow reservoir regime [see Figs.~\ref{figure2}(d3) and \ref{figure2}(d4)] and smaller ellipticity for small $s_z$ [see Figs.~\ref{figure2}(d2) and \ref{figure2}(d4)].

\begin{figure*}
  \includegraphics[width=1.0\textwidth]{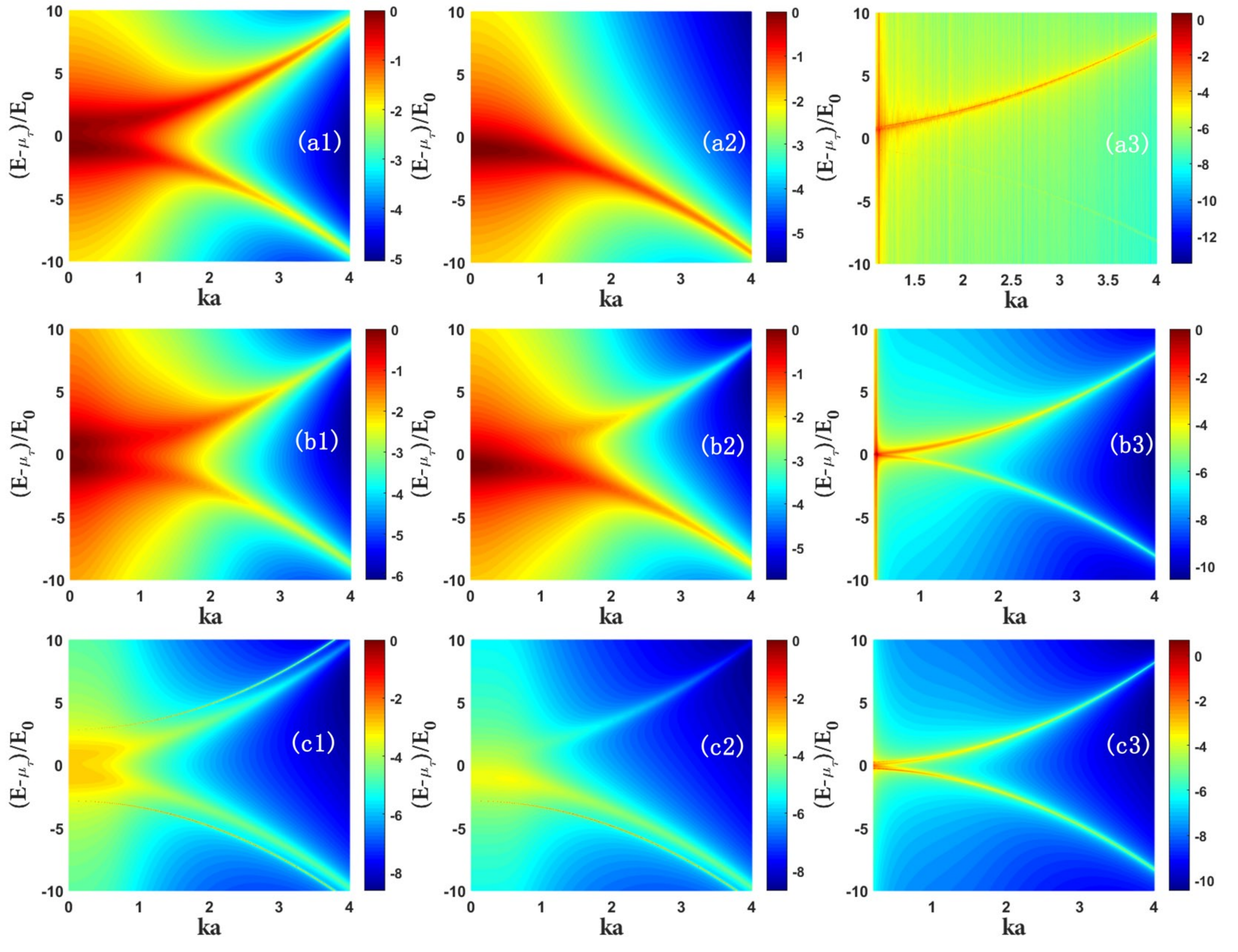}\\
\caption{(Color online) PL spectrum  $\log_{10}(PL/PL_{max})$ [see Eq.~(\ref{PL})] of a spinor polariton condensate for different parameters. Panels (a)-(b): PL spectrum of a unpolarized spinor polariton condensate with (a) $g_{12}>0$, (b) $g_{12}<0$, and $\Omega/E_0=0.1$. Panel (c): PL spectrum of a polarized spinor polariton condensate. For other parameters, we take (a1) $g_{12}/g=0.3$, $\gamma_R/\gamma_C=0.1$,  $k_BT/E_0=2$;  (a2) $g_{12}/g=0.3$, $\gamma_R/\gamma_C=0.1$, and $k_BT/E_0=0.1$; (a3) $g_{12}/g=0.3$, $\gamma_R/\gamma_C=1.8$, and $k_bT/E_0=2$; (b1) $g_{12}/g=-0.3$, $\gamma_R/\gamma_C=0.1$, and $k_BT/E_0=2$; (b2) $g_{12}/g=-0.3$, $\gamma_R/\gamma_C=0.1$, and $k_BT/E_0=0.1$; (b3) $ g_{12}/g=-0.3$, $\gamma_R/\gamma_C=1.8$, and $k_bT/E_0=0.1$; (c1) $g_{12}/g=2.5$, $\gamma_R/\gamma_C=0.1$, and $k_BT/E_0=2$;  (c2) $g_{12}/g=2.5$, $\gamma_R/\gamma_C=0.1$, and $k_BT/E_0=0.1$;  (c3) $g_{12}/g=2.5$, $\gamma_R/\gamma_C=1.8$, and $k_BT/E_0=0.1$. In all plots, the length, the energy, and the time are scaled in units of $a=\sqrt{\hbar/m\gamma_C}$,  $E_0=\hbar \gamma_C$ and $\tau_c=\hbar/E_0$ with $a$ being the experimental length scale (e.g., in GaAs $a=2$ $\mu$m, $\tau_c=3$ ps, and $E_0=0.68$ meV). }\label{figure3}
\end{figure*}

\section{Photoluminescence of a Spinor Polariton Condensate}\label{section:four}

In this section, we discuss how to experimentally probe the various Bogoliubov dispersions of a spinor polariton condensate presented in this work. Recently, significant experimental progress has been achieved on measuring elementary excitations of a one-component polariton condensate, aiming particularly at observing two features: (1) the diffusive modes, which originate from the driven-dissipative character of the system \cite{Wouters2007,Oexcitation0}; (2) the negative-energy ghost branch (GB) of the Bogoliubov dispersion, which arises from the hole component of excitations and thus mirror images the normal positive-energy branch (NB). In experiments reported in Refs.~\cite{Excitation1,Excitation2}, while NB has been directly observed in the photoluminescence (PL) of a nonresonant polariton condensate  \cite{ExcitationE1,ExcitationE2}, no fingerprints of GB were spotted. Since the GB signal may be easily masked by strong emission, whether it is possible to detect GB in a non-resonant polariton BEC was subjected to debate. On the other hand, the GB dispersion was detected in a resonantly pumped polariton condensate using the four-wave mixing technique, as reported in Ref. \cite{FSM}. Remarkably, the first successful observation of the PL signal reflecting GB in a non-resonant polariton condensate was reported in Ref. \cite{PL1}. Motivated by these experimental advances, below we are interested in studying the PL spectra of a spinor polariton BEC under non-resonant pumping, in particular, the visibility of the ghost branches of dispersions of both the global- and relative-phase modes, extending relevant work on one-component polariton BECs \cite{Byrnes2012,PL0}.

At the heart of the measurement of the excitation spectrum with PL is the measurement of two-time correlation function of the spinor polariton condensate. Denoting the PL spectrum by $\textrm{PL}\left(k,\omega\right)$, we exploit the approach in Refs. \cite{Byrnes2012,App3} and calculate as follows. Defining a matrix $V^{-1}$ in the form
\begin{eqnarray}
V^{-1}=\left(\begin{array}{ccccc}
\overline{u}_{11} & \overline{u}_{12} & \overline{u}_{13} & \overline{u}_{14} & \overline{u}_{15}\\
\overline{u}_{21} & \overline{u}_{22} & \overline{u}_{23} & \overline{u}_{24} & \overline{u}_{25}\\
\overline{u}_{31} & \overline{u}_{32} & \overline{u}_{33} & \overline{u}_{34} & \overline{u}_{35}\\
\overline{u}_{41} & \overline{u}_{42} & \overline{u}_{43} & \overline{u}_{44} & \overline{u}_{45}\\
\overline{u}_{51} & \overline{u}_{52} & \overline{u}_{53} & \overline{u}_{54} & \overline{u}_{55}
\end{array}\right),
\end{eqnarray}
we diagonalize the Bogoliubov's matrix [see Eq.~(\ref{BogMatrix})] as $V^{-1}\mathcal{L}_{{\bf k}}V=(E_1,-E_1^*,E_1,-E_2^*,E_3)$. Here, the eigenvalues $E_1$, $E_2$, and $E_3$ correspond to the
density mode, spin-density mode, and reservoir mode, respectively. The PL spectrum can then be derived as
\begin{widetext}
\begin{eqnarray}
\textrm{PL}\left(k,\omega\right)&\propto&\left|C_{k}\right|^{2}\text{Re}\Big\{ \frac{i n_{1k}\left(\overline{u}_{11}\overline{u}_{22}+\overline{u}_{31}\overline{u}_{42}\right)}{\hbar\omega-E_{1}}+\frac{i \left(n_{1k}+1\right)\left(\overline{u}_{12}\overline{u}_{21}+\overline{u}_{32}\overline{u}_{41}\right)}{\hbar\omega+E_{1}^{*}}+\frac{i n_{2k}\left(\overline{u}_{13}\overline{u}_{24}+\overline{u}_{33}\overline{u}_{44}\right)}{\hbar\omega-E_{2}}\nonumber\\
&+&\frac{i\left(n_{2k}+1\right)\left(\overline{u}_{14}\overline{u}_{23}+\overline{u}_{34}\overline{u}_{43}\right)}{\hbar\omega+E_{2}^{*}}\Big\}, \label{PL}
\end{eqnarray}
\end{widetext}
where $C_k$ is called the Hopfield coefficient for the photonic component of the polaritons, and $n_{1k}$ ($n_{2k}$) is the thermal population of quasiparticles associated with the density mode (spin mode). Different from the PL spectrum of a one-component poariton condensate, Eq.~(\ref{PL}) involves contributions from both the global- and relative-phase excitations: the first (last) two terms correspond to the global-phase (relative-phase) mode, with the negative branch contained in the second (fourth) term, respectively. In Fig.~\ref{figure3}, we present the PL spectrum of a spinor polariton condensate for various parameters.

Let us first compare the PL spectrum in the spin-unpolarized phase and the spin-polarized phase considering, for example, the parameter regime $\gamma_R\ll \gamma_C$ and high temperature (larger than the relevant energy gap). As illustrated in Figs.~\ref{figure3}(a1) and \ref{figure3}(c1), both plots show the positive and negative branches of the dispersions, due to large populations at high temperatures. Yet, a prominent feature of Fig.~\ref{figure3}(c1) for the spin-polarized phase, as compared to Fig.~\ref{figure3}(a1) for the spin-unpolarized phase, is the appearance of two separated sectors of dispersions, both in the positive and negative branches. This second dispersion sector, with a very narrow linewidth, corresponds to the spin mode which is weakly damped in the spin-polarized phase. Instead, the dispersion of the relative-phase mode does not appear in the spin-unpolarized phase [see Fig.~\ref{figure3}(a1)], as the relative-phase mode there only has real energy and is undamped [see Fig.~\ref{figure1}(b)]. Thus, for the spin-unpolarized phase in the present setup, only the global-phase excitation sector of dispersions are revealed by PL spectrum, [see Fig. \ref{figure3}(a) and \ref{figure3}(b) for various parameters]. As illustrated in Fig. \ref{figure3}(c2), at low temperatures, only the negative-energy ghost branch of the relative-phase mode becomes visible, due to strongly suppressed thermal population of the gapped positive-energy relative-phase mode. In contrast, the relative-phase excitation sector cannot be distinguished in Fig.~\ref{figure3}(c3) corresponding to a fast reservoir regime. This can be attributed to the fact that, for this particular parameter choice, the real part of the spectra of the global-phase mode is very close to the relative-phase mode [see Fig.~\ref{figure2}(a2)]. To conclude, the negative-energy ghost branch of the relative-phase mode can be clearly distinguished from that of the global-phase mode in the PL spectra of a spinor polariton condensate in the spin-polarized phase, where the relative-phase mode is weakly damped. Its resolution is optimal when the (real part) energy of the relative-phase mode is sufficiently separated from the global-phase mode, such that it is still resolvable after taking into account of the finite linewidth of the spectrum of the global-phase mode.

Turning back to the spin-unpolarized phase, as discussed above, here only dispersions of the global-phase mode can be visualized in the PL spectrum. In general, the width of the spectrum exhibits a significant broadening for $\gamma_R\ll \gamma_C$ [see, e.g., Fig.~\ref{figure3}(a1)] reflecting a strong damping of the global-phase mode [see also Fig.~\ref{figure1}(b3)]. Instead, a much narrower spectrum is shown in the fast reservoir limit [see Fig.~\ref{figure3}(a3)], where the global-phase mode only weakly decays. Notice that in Fig.~\ref{figure3}(a3), despite high temperature, no fingerprints of the negative-energy dispersion of global-phase mode is observed. This is due to the fact that the corresponding Bogoliubov coefficients in Eq.~(\ref{PL}) nearly vanish for this specific parameter choice. Considering experimental relevant parameters, we plot the corresponding PL spectrum in the spin-unpolarized phase in Figs.~\ref{figure3}(b). Compared to Figs.~\ref{figure3}(a), we see that the sign of $g_{12}$, does not qualitatively change the features of the PL spectrum in the spin-unpolarized phase.

\section{Conclusion}\label{section:five}

Summarizing, we have theoretically studied the steady phases and elementary excitations of a spinor polariton condensate created by non-resonant excitation, assuming that $g_{12}/g$ can be widely tuned, and fast spin-relaxation in reservoir. In the regime $g_{12}<g+2\Omega/n_0$, the polariton condensate is in the spin-polarized phase, exhibiting a linear polarization whose direction is pinned by the $\Omega$ term. In the regime $g_{12}>g+2\Omega/n_0$, the condensate is in the spin-polarized phase, exhibiting an elliptical polarization. The transition occurs at the critical interaction, where the spin-density response function diverges and the energy gap of the relative-phase mode closes. We have compared behavior of elementary excitations in two phases, taking into account  reservoir effects in the crossover from $\gamma_R\ll \gamma_C$ to the limit of $\gamma_R\gg \gamma_C$.  We have shown that the gapped relative-phase modes are long lived in both phases and are robust to reservoir effects. The energy spectrum of the global-phase modes, by contrast, are strongly tailored by the reservoir effect, being gapped in the slow reservoir limit but diffusive in the opposite fast reservoir regime. In the spin-unpolarized phase, the mode polarization is always linear, one copolarizing with the condensate and the other cross polarizing with it. However, in the spin-polarized phase, the reservoir effect has a particular strong impact on polarization of the global-phase mode at small momenta. While the polarization of relative-phase mode is weakly influenced by reservoir, it is sensitive to the circular polarization degree of the condensate. In addition, we have demonstrated that the energy dispersions presented in this work can be directly observed in the PL emission. In particular, we show that the negative-energy ghost branch of the dispersion of the relative-phase mode can be clearly visualized in the spin-polarized phase, exhibiting a very narrow linewidth and distinguishable from that of global-phase mode. That the relative-phase mode is undamped in the spin-unpolarized phase, leading to its absence in the corresponding PL spectrum, may be related to the fact that in our model the reservoir excitons are assumed to be coupled to the total density, rather than the spin density, of the spinor polariton condensate [see Eq.~(\ref{Rate})]. Hence, an interesting subject of investigation in the future consists of cases with asymmetric couplings to reservoir polaritons or asymmetric decay rates of condensate polaritons.

\begin{acknowledgments}
We thank Y. Xue, C. Gao, and B. Wu for stimulating discussions. This is supported by the NSFC of China (Grants No. 11274315 and No. 11374125) and Youth Innovation Promotion Association CAS (Grant No. 2013125). Y. H. acknowledges support from Changjiang Scholars and Innovative Research Team in University of Ministry of Education of China and PCSIRT (Grant No. IRT13076) the NSFC of China (Grant No. 11434007). Z. D. Z. is supported by the NSFC of China (Grant No. 51331006).
\end{acknowledgments}

\appendix
\section{Symmetry analysis of Bogoliubov matrix (\ref{BogMatrix1})}\label{AppendixA}

In this appendix, we analyze the symmetry properties of Bogoliubov matrix $\mathcal{L}_k$ in Eq.~(\ref{BogMatrix1}) in the linearly polarized case. We notice that $\mathcal{L}_k$ is invariant under the following two transformations:
\begin{eqnarray}
U_{1}	=\left(\begin{array}{ccccc}
0 & 0 & 1 & 0 & 0\\
0 & 0 & 0 & 1 & 0\\
1 & 0 & 0 & 0 & 0\\
0 & 1 & 0 & 0 & 0\\
0 & 0 & 0 & 0 & 1
\end{array}\right),\
U_{2}	=\left(\begin{array}{ccccc}
0 & 0 & 1 & 0 & 0\\
0 & 0 & 0 & -1 & 0\\
1 & 0 & 0 & 0 & 0\\
0 & -1 & 0 & 0 & 0\\
0 & 0 & 0 & 0 & 1
\end{array}\right).
\end{eqnarray}
The consequence of this symmetry property is, if $V=\left(\begin{array}{ccccc}
u_{1{\bf k}} & v_{1{\bf k}} & u_{2{\bf k}} & v_{2{\bf k}} & w_{{\bf k}}\end{array}\right)^{T}$ is an eigenvector of Bogoliubov matrix (\ref{BogMatrix1}), the action of $U_{1}$ ($U_{2}$) on $V$ realizes a simultaneous exchange: $u_{1{\bf k}}\leftrightarrow u_{2{\bf k}}$ and $v_{1{\bf k}}\leftrightarrow v_{2{\bf k}} $  ($u_{1{\bf k}}\leftrightarrow-u_{2{\bf k}}$ and $v_{1{\bf k}}\leftrightarrow-v_{2{\bf k}}$ ), such that $U_{1}V$ ($U_{2}$V) is also an eigenvector of $\mathcal{L}_k$ with the same eigen-value.

\bibliography{myr}
\end{document}